\documentclass[manuscript]{acmart} 

\AtBeginDocument{%
  }

\setcopyright{acmlicensed}
\copyrightyear{2025}
\acmYear{2025}
\acmDOI{XXXXXXX.XXXXXXX}

\begin{document}

\title{The Turn to Practice in Design Ethics: Characteristics and Future Research Directions for HCI Research}

\author{Gizem Öz}
\email{gizemoz@cc.au.dk}
\orcid{0000-0001-5408-1772}
\affiliation{%
  \institution{Aarhus University}
  \city{Aarhus}
  \country{Denmark}
}

\author{Christian Dindler}
\email{dindler@cavi.au.dk}
\orcid{0000-0002-4914-3323}
\affiliation{%
  \institution{Aarhus University}
  \city{Aarhus}
  \country{Denmark}
}

\author{Sharon Lindberg}
\email{sharon@dsv.su.se}
\orcid{0000-0003-4084-3259}
\affiliation{%
  \institution{Stockholm University}
  \city{Kista}
  \country{Sweden}
}

\renewcommand{\shortauthors}{Öz et al.}

\begin{abstract}
As emerging technologies continue to shape society, there is a growing emphasis on the need to engage with design ethics as it unfolds in practice to better capture the complexities of ethical considerations embedded in day-to-day work. Positioned within the broader "turn to practice" in HCI, the review characterizes this body of work in terms of its motivations, conceptual frameworks, methodologies, and contributions across a range of design disciplines and academic databases. The findings reveal a shift away from static and abstract ethical frameworks toward an understanding of ethics as an evolving, situated, and inherent aspect of design activities, one that can be cultivated and fostered collaboratively. This review proposes six future directions for establishing common research priorities and fostering the field’s growth. While the review promotes cross-disciplinary dialogue, we argue that HCI research, given its cumulative experience with practice-oriented research, is well-equipped to guide this emerging strand of work on design ethics.
\end{abstract}



\received{29 May 2025}
\received[revised]{}
\received[accepted]{}


\maketitle

\begin{text}
{Author’s Original Manuscript} 
\end{text}

\section{Introduction}

The ethics of design in Human-Computer Interaction (HCI) has been a topic of research for decades \cite{shilton2018values}, and questions related to the responsibilities and moral obligations of designers appear in well-established disciplines such as participatory design \cite{bodker2022participatory} and value-sensitive design \cite{friedman2013value}, as well as in recent work on sustainability \cite{hansson2021decade}, designing for more-than-human futures \cite{eriksson2024more} and designing with care \cite{garrett_felt_2023}. Despite this long-standing engagement with ethics, several authors have noted that much research in HCI and design has treated ethics as an a priori and static concept \cite{frauenberger_-action_2016}, with less attention paid to how ethical issues unfold in design practice \cite{gray_ethical_2019}. However, recent years have seen growing interest in considering ethics from the perspective of practice, specifically, how designers, technology developers, and the organizations they work within approach ethical issues in their everyday work. Rooted in the broader turn to practice in HCI \cite{bodker_when_2006, kuutti_turn_2014}, which emphasizes the complex and situated nature of everyday interactions, much of the research on design ethics in practice has emerged from the HCI research. This body of work highlights the increasing ethical implications of design, particularly in the context of emerging technologies and their widespread impact on daily life.

Despite a growing number of papers focusing on empirical studies of design ethics in practice \cite{lindberg_design_2021, dindler_engagements_2022, gray_ethical_2019}, the research remains nascent, and contributions are scattered across HCI and design research venues. In this paper, we aim to take stock of this emerging field by characterizing what we describe as a "turn to practice in design ethics research". We explore the motivations, methods, theories, and key findings of research, and we outline a range of gaps and potentials for developing future research directions to advance this turn in design ethics research. While previous reviews have addressed ethical issues related to AI and algorithms \cite{berman_scoping_2024, hajigholam_saryazdi_algorithm_2024}, specific user groups such as children \cite{van_mechelen_18_2020}, specific uses of the concept of ethics in SIGHCI \cite{nunes_vilaza_scoping_2022}, or specific ways of working \cite{boyd_adapting_2021}, this paper offers a review of contributions across HCI and design that approach ethics from the perspective of practice. Thus, this scoping review is guided by the following research questions:

\begin{itemize}

\item What is state-of-the-art in studying ethical engagements within design practices? 

\item What characteristics does the literature exhibit in terms of theories, concepts, and methodologies employed to explore design ethics in practice?

\item Which gaps and potential future directions can be identified and grounded within HCI research? 
\end{itemize}

In HCI research, design practice has drawn growing attention since the "turn to design" in the 1990s and the "turn to practice" in the 2010s, leading to a deeper focus on the complex, situated nature of design work  \cite{gray_dark_2018}. In particular, UX-related design work has been recognized as a distinct and valuable form of inquiry, deeply rooted in context, guided by professional expertise and ethical responsibility, and following its own standards of rigor (e.g.,  \cite{sondergaard_troubling_2020, inie_how_2020, lee_purpose_2025}). The HCI field is actively engaging in discussions about how design practice and its tools relate to academic research, its ethical dimensions, and how theoretical approaches such as practice theory might contribute to advancing design within the HCI field \cite{gray_ethical_2019, pink_applying_2013, gray_using_2022, lim_anatomy_2008}.

Our definition and boundaries of design practice for this review draw heavily on the conceptual frameworks offered by Stolterman \cite{stolterman_nature_nodate} and by Wolf et al.´s \cite{wolf_dispelling_2006} articulation of Löwgren's \cite{lowgren_applying_1995} creative design. Löwgren \cite{lowgren_applying_1995} describes creative design work as an ongoing dialogue between identifying the problem and finding solutions, emphasizing understanding the problem just as much as developing the final product. This process often involves exploring a wide range of ideas and re-evaluating initial assumptions, making it inherently uncertain and personal to the designer. Within the scope of this paper, design practice refers to creative design practices, including UX design, as well as design work that complements or intersects with HCI, such as graphic design, industrial design, and visual communication design. While our primary aim is to contribute to the field of design in HCI research, we argue that looking across a broader range of design fields and converging HCI and design literature is essential for building a more comprehensive understanding of ethics in practice. Many of the challenges faced in design work in HCI, such as dealing with ambiguity, balancing competing values, and navigating real-world constraints, are not unique to HCI but are shared across design disciplines. This cross-disciplinary view strengthens our ability to address the complex, situated nature of ethical challenges. Our goal is to consolidate these diverse perspectives into HCI to better address the complexities of design culture and practice within the context of technology development. This focus is reflected in the review method, where the search strategy was developed to identify papers across both HCI and design research venues. A cross-field review also demonstrated that, given its cumulative experience with practice-oriented research, the HCI community is well-positioned to lead the development of this emerging area of design ethics.

This paper offers several contributions to advancing design ethics research within HCI. First, we characterize the turn to practice in design ethics research by pinpointing shared motivations, conceptualizations, methods, and contributions. In doing so, we aim to meaningfully bring together these diverse works under the concept of the turn to practice in HCI, and to promote a shared understanding of design ethics in practice that supports a more unified research landscape. Second, by synthesizing studies scattered across various venues and design disciplines, we seek to foster a more cohesive and dialogic research community. Third, drawing on our findings, as well as the gaps and potentials identified in the current literature, we propose several concrete directions for strengthening and expanding this area of research.

The article is organized in the following way. Section 2, Background, begins with an overview of how the concept of practice has been understood and studied across different paradigms within HCI. It then focuses on approaches from the third paradigm HCI and the turn to practice in HCI, highlighting their relevance for research in design ethics. Section 3, Review Process, details how the literature was collected, analyzed, and synthesized. Section 4, The Results, and Section 5, Analysis,  present our key findings, including: (i) the motivations driving the studies and how these informed practice-oriented research approaches; (ii) the common conceptualizations used, including shared attributes of how the concept of ethics is framed across studies; (iii) the methods employed and how they address the situated, everyday nature of ethics; and (iv) the overarching research aims, findings, and contributions, with links drawn across these dimensions. Section 6, The Discussion, characterizes the turn to practice in design ethics as a research strand bringing together the main insights and introduces six key directions for future work. The final section concludes by summarizing the key contributions of the review.

\section{Background}

\subsection{The Turn to Practice in HCI Research}

The history of HCI is often described through a series of waves or paradigms, each representing a shift in focus, theoretical foundations, and methodologies \cite{rogers_hci_2012, bodker_third-wave_2015, harrison_three}. Across these waves, the concept of "practice"—both in terms of its meaning and how it is studied— has evolved significantly. In first wave HCI, the practice in focus was a scientific and engineering-based approach to designing and evaluating human-computer interaction \cite{bodker_third-wave_2015}. It emphasized optimizing the fit between humans and machines, mainly in work settings, using formal methods, cognitive theories, and performance metrics. Design followed analysis and aimed to improve usability by making systems more efficient, error-free, and easy to learn. It can be said that, in this view, social or cultural aspects of technology use were not central \cite{bodker2022participatory, bannon_humanfactors, harrison_three}. 

In second wave HCI, practice in focus shifted from individual users to understanding how groups use technology in real work settings \cite{rogers_hci_2012, bodker_when_2006}. It emphasized social interaction, context, and collaboration within communities of practice. Theoretical frameworks such as situated action, distributed cognition, and activity theory became central \cite{rogers_hci_2012}. These approaches offered new ways to understand how technology supported or disrupted the flow of everyday work. Design aimed not only for usability but also for supporting group work and addressing social and organizational issues, laying the foundation for later shifts toward understanding technology as part of broader social practices. The methods used in this wave moved away from rigid, formal testing toward more flexible and exploratory approaches such as participatory design and ethnographic fieldwork, prototyping, and contextual inquiries\cite{bodker_third-wave_2015, bodker_when_2006, fallman_new_2011}. Although the second wave did not fully embrace the practice turn that would define later developments in HCI, it laid important groundwork. The growing interest in context, social interaction, and collective activity helped prepare the field for the more fundamental shifts that would emerge in the third wave \cite{kuutti_turn_2014}.

The third wave \cite{bodker_when_2006}   \cite{harrison_three} extended HCI´s focus beyond the workplace into areas such as leisure, home, and public spaces. It represents a significant shift, emphasizing interaction as situated within the social and embodied complexities of everyday life \cite{frauenberger_-action_2016, kim_invest, judge_sharing_2010}. Rather than targeting clearly defined tasks or user groups, it explores how people interact with technology in diverse and often ambiguous ways. It highlights situatedness \cite{frauenberger_-action_2016, kuutti_turn_2014}, meaning-making \cite{pink_applying_2013}, embodiment \cite{frauenberger_entanglement_2020}, and experience \cite{kuutti_turn_2014}—understanding interactions as embedded in specific social, cultural, and physical contexts rather than isolated events. As an example of this move toward everyday settings, Rogers and Marshall \cite{rogers_wild} propose "research in the wild" as an umbrella term for approaches that study technology in natural, real-world environments rather than controlled lab settings. They defined this term as an effort to understand how people use and interact with technology in everyday life, using a combination of qualitative methods such as interviews and observations, as well as quantitative data, with a focus on how technology integrates into daily routines \cite{rogers_wild}.

The concept of practice takes on a deeper and more central meaning in the third wave than in earlier paradigms. The third wave—and what is often referred to as the turn to practice—repositions practice as both the unit of analysis and the unit of design. It aims to understand the complex and situated nature of technologies embedded in seemingly mundane activities and routines \cite{kuutti_turn_2014, entwistle_beyond_2015, kuijer_practices_2013}. Kuutti and Bannon \cite{kuutti_turn_2014} advocate for a practice-oriented approach, encouraging research that addresses real-world challenges encountered by individuals, communities, or industries rather than centering on abstract or purely theoretical concerns. This turn to practice has been particularly influential in areas such as sustainable HCI, where the focus has shifted from changing individual behaviors to understanding and intervening in energy-consuming practices (e.g., the 2013 TOCHI special issue on "Practice-Oriented Approaches to Sustainable HCI" \cite{pierce_introduction_2013}). This shift draws heavily on social science theories of practice from fields such as anthropology, sociology, and cultural studies \cite{pierce_introduction_2013}. For example, Reckwitz's \cite{reckwitz_theory_2002} elements of practice (bodily and mental activities, things and their use, background knowledge, emotions, and motivation) and Shove et al.'s \cite{shove_dynamics_2025} framework (practice as a combination of materials, competences, and meanings) have informed HCI approaches to studying the organization and reorganization of shared activities and routines \cite{kuijer_practices_2013,lindberg_cultivating_2020, entwistle_beyond_2015, kuutti_turn_2014, soden_climate_2025, tomlinson_collapse_2013}. These frameworks provide empirical perspectives that consider the historical, structural, and contextual dimensions of practice, as well as interconnections with other practices \cite{kuutti_turn_2014}. 

Within these theories, the concept of practice has been interpreted in various ways. It can be seen as the implicit knowledge or skills that guide human actions, or as organized sets of activities involving both people and material artifacts, or as sociomaterial arrangements closely linked to specific contexts, individuals, and interactions \cite{guzman_grey_2013}. In any case, two common threads across these interpretations are: (1) practice is the setting where actions, decisions, and negotiations take place, and (2) practices are shaped by and embedded within broader technological, institutional, and infrastructural contexts \cite{pink_applying_2013, entwistle_beyond_2015}. Wearn and Back \cite{wearn_activity} suggest understanding design knowledge as "activity"—that is, knowledge in design is not something fixed or separate from the act of designing, but rather something that emerges through the ongoing process itself. They develop a practice-oriented view of design knowledge by emphasizing that it is generated through the act of designing, where materials, ideas, concepts, and human actors co-evolve. Rogers \cite{rogers_new_theo} defines "the practitioner’s perspective" in research in the wild as accounts from "people who work in industry and are in the business of researching, designing, and evaluating products" (p.24) In broad terms, design practice refers to the contexts in which activities such as ideation, sketching, and prototyping are carried out, as well as where ethics-related decisions are made and embedded.

Amid the growing interest in ethics across HCI and design, a recent strand of work studies ethics as it unfolds in design practice to understand the complex, context-specific nature of design ethics—and the active role designers play in recognizing and engaging with ethical challenges. Upcoming sections of this paper review and situate this body of work within the broader turn to practice in HCI research, linking it to the epistemological foundations that characterize third wave HCI research.

\subsection{The Turn to Practice in Design Ethics in HCI Research}

The question of ethics runs throughout the history of HCI and design \cite{shilton2018values, chan_design_2018}, and has been regularly rearticulated in response to major developments in architecture, product design, and, more recently, the widespread adoption of digital technologies—most notably, artificial intelligence. Ethics has long played a central role in traditional disciplines such as architecture, particularly in debates over how the built environment enables or impedes the good life, as seen in movements like the Bauhaus \cite{ramos_2019}. In product design, ethical concerns have likewise featured prominently in ongoing discussions around sustainability, production methods, and the viability of circular economies. From an academic perspective, leading design journals have often explored the need for "discipline-specific ethics" within design \cite{donahue2004discipline, jonas2006special}, as well as how classical ethical theories such as deontology and consequentialism might provide a theoretical foundation for design ethics \cite{mitcham1995ethics}. Additionally, scholars have drawn on ethical traditions from thinkers such as Aristotle \cite{bousbaci2005more}, Levinas, and Derrida \cite{steen2012human} to capture the nuanced nature of ethics in design. Despite this seemingly rich discourse, there remains skepticism about the field’s tangible progress in integrating ethics into practice \cite{fry2004voice}, prompting calls for a renewed focus on ethics in design \cite{chan_design_2018}.

In HCI, ethical considerations can arguably be traced to the very origins of the field, particularly through ongoing discussions about technology-mediated human activity; evolving from early critiques to developing generative frameworks and focusing on how ethics plays out in real-world, situated contexts \cite{shilton2018values}. Early scholars like Wiener contributed significantly to computer ethics, exploring issues such as privacy, surveillance, fairness, and accountability \cite{dindler_engagements_2022, shilton2018values}. Throughout the 1990s, within HCI, the CSCW scholarship was founded on the value of cooperation and addressed issues like empowerment and privacy. PD similarly engaged with ethical issues such as cultural norms and legitimate participation \cite{shilton2018values}. Fallman \cite{fallman_new_2011} argues that the rise of user-centered design —strongly influenced by Scandinavian traditions— brought ethical and value-based concerns to the forefront of HCI. This included involving users early in the design process and considering their needs, values, and concerns, especially regarding power, labor, and workplace politics. According to Shilton\cite{shilton2018values}, earlier work was mainly critical, analyzing the unintended consequences of technologies and emphasizing perspectives from feminist, anti-racist, and postcolonial critiques. These analyses shaped the field's understanding of values and ethics, helping to inform future work. As they went along, researchers began developing generative approaches that aimed to integrate ethical thinking into the design process itself, and ethical issues have been evident in various design approaches \cite{shilton2018values, steen2015upon}.

Within third wave HCI, one of the most well-known frameworks is Value Sensitive Design (VSD), which combines theoretical, empirical, and technical methods to consider the needs of stakeholders \cite{shilton2018values}. VSD encourages designers to think about both direct and indirect stakeholders using tools like stakeholder analysis, scenarios, and interviews. Despite its contributions, VSD has also been critiqued for potentially favoring the values of researchers over those of participants, particularly when relying on predefined heuristics rather than more context-sensitive, participatory methods \cite{fallman_new_2011, shilton2018values}. Other generative methods have emerged to complement or expand upon VSD. For example, "Values at Play" and "worth-centered design" bring more artistic and creative interpretations to ethical design, focusing less on rationality and more on experiential worth \cite{shilton2018values}. Similarly, approaches like critical design and making take a broader perspective by questioning the direction and purpose of technology itself; focusing on uncovering and addressing hidden values within the design process, aiming to challenge limitations and give voice to perspectives that are often overlooked. Another significant body of work addresses the notion of ethics as embedded in digital artifacts—mediating actions and promoting certain values or perceptions over others \cite{verbeek_materializing_2006}.

Ethics research has now been further fueled by public debates around data, privacy, and the use of AI, spurring research on issues such as dark patterns in user interfaces \cite{gray2023mapping, lee_purpose_2025, kraus_what_2024} and AI ethics \cite{khan2022ethics}. The current state of HCI  and design researchers grappling with ethical practice is marked by a growing realisation of technology's significant impact on culture and a corresponding concern for the responsibility of designers. A more recent direction in the field argues for the need to study ethics as it is encountered within design practice. Traditional, anticipatory, and formalised ethics processes, often rooted in research models and focused on avoiding risk and harm, are seen as insufficient to manage the ethical dimensions that emerge in the situated, value-driven, and participatory nature of third-paradigm HCI \cite{frauenberger_-action_2016}. The way in which work and practice are conceptualized directly shapes how ethics is perceived and enacted. Using Gherardi’s framing \cite{gherardi_introduction_2009}, practice-oriented research in design ethics examines how practitioners engage with ethical questions through their work—how they think, feel, and talk about ethics—is affected by both "outside" sources such as organizational structures or technologies (e.g., \cite{gray_building_2024}), or from "inside" sources like personal skills or emotions (e.g., \cite{popova_vulnerability_2022, garrett_felt_2023}). Since ethical questions often emerge from within practice itself, a full understanding requires a holistic view that considers the many intertwined elements shaping values and behavior \cite{dindler_engagements_2022, pillai_exploring_2022, gray_ethical_2019, chivukula_dimensions_2020}.

Among the main arguments for this turn to practice is the need (1) to understand ethics as an emergent phenomenon \cite{frauenberger_-action_2016},(2)  to understand ethical issues "on their own terms" \cite{chivukula_dimensions_2020} within practical and organizational complexity, (3) to address the situational nature of ethics \cite{munteanu2015situational} and (4) to eventually build concept and methods with "practice resonance" \cite{gray_scaffolding_2023}. Such arguments have motivated research into how design practitioners perceive, articulate, and negotiate ethics \cite{gray_languaging_2024, dindler_engagements_2022, gray_ethical_2019, rivard_articulating_2021}, the material and organizational circumstances in which ethical questions arise \cite{boyd_adapting_2021, chivukula_dimensions_2020, watkins2020tensions}, and how ethics might be collaboratively cultivated \cite{lindberg_cultivating_2020}. This body of work also reflects on the methodological challenges of studying ethics in practice \cite{boyd_adapting_2021}. 

In sum, the turn to practice in HCI and design research marked a shift from focusing on individual cognition and isolated artifacts to examining how knowledge, meaning, and design work emerge through everyday activities and social interactions. In HCI ethics research, this shift has emphasized that ethical issues are not merely theoretical concerns but are shaped by practitioners’ situated actions, negotiations, and collective sense-making within real-world work contexts.  It is timely to bring together these efforts, underline their embeddedness in the turn to practice in HCI, and reflect on their key features to help establish a more coherent and unified area of research. 

\section{Review Process}

Our research questions and review protocol are formulated to map the breadth of the research field, suitable for the frame of the scoping review \cite{pare_synthesizing_2015, arksey_scoping_2005}. We look for work that reports on empirical studies involving practitioners through methods such as observation, interviews, or collaborative sessions, and work that explicitly addresses ethics as it relates to design practice. To develop a comprehensive characterization of design ethics in practice and its emerging variants, we conducted literature searches across three major databases: Web of Science (WOS), the Association for Computing Machinery (ACM) Digital Library, and the Design Research Society (DRS) Digital Library. WOS was chosen for its extensive and diverse journal coverage. Our preliminary research revealed several papers on design ethics and practice within the ACM and DRS Digital Library, and it was thus also selected as a key database to include. We included full-length research articles published in English in journals and conferences indexed in these databases over the past ten years; abstracts, pictorials, editorials, and short papers were excluded.

\subsection{Identifying Relevant Literature: Information Sources and Search Queries}
To create a wide but focused set of search queries, we broke down our research questions into keyword categories and searched combinations of these keywords to cover the area. In addition to the keywords "design*" and "ethic*", we used "design team", "designer", and "practitioner" to cover the context of the practice component of the research. To specify the empirical methods, we used the keywords: "interview", "workshop", "observation", "qualitative", "case study", "fieldwork", "codesign" and "participatory design". We searched ethics and practice-related keywords in the "Abstract," and method-related keywords in "All fields," as sometimes articles do not specify their methods in the abstract. Searching with combinations of these terms generated a total of 24 search queries. The search was executed between September and November 2024.

We narrowed down the results with the publication filter: any general HCI and design-related journal and conference proceedings, as well as more scope-specific journals such as social design, visual design, and game design, are included. After completing all the queries, we checked whether a set of articles that we had handpicked and knew beforehand were included as validation of our search set.

\subsection{Selecting the Literature: Inclusion and Exclusion Criteria}
We conducted an initial screening of the abstracts of the papers identified during the search process. We spent a great deal of time establishing thorough inclusion and exclusion criteria because this is essential to structuring a consistent scoping review \cite{pare_synthesizing_2015}. To refine and finalize our inclusion criteria, article-by-article deliberations were conducted on selected papers exemplifying boundary cases. This led to the following: 

\begin{itemize}
    \item  Papers were included only if their central focus was design ethics in practice; articles that merely mentioned ethics in passing were excluded.

    \item  We focused on papers that generated insights through empirical research methods, such as interviews, observations, or case studies within design practices. Theoretical and review papers, as well as those analyzing materials such as design manifestos \cite{fritsch_calling_2018}, job listings \cite{rismani_what_2023}, or Reddit forums \cite{gray_what_2020}, were excluded, as they did not provide direct insights from design practice.

    \item  The review intends to characterize a movement in design ethics research that we frame as a turn to practice. As such, to focus our publication pool, we omitted papers focusing on specific moral themes such as sustainability, vulnerable users and communities, inclusivity, and accessibility, or privacy and transparency. Our search queries caught these papers because they used the "ethic*" keyword in their abstracts. Although these themes are related to ethics in general, including them would have expanded our scope to an unmanageable range of moral topics. We thus excluded these papers to maintain our focus on design ethics in practice and to keep the number of papers manageable. We employed the same criteria for omitting papers using themes adjacent to ethics, such as responsible innovation, and corporate social responsibility, as well as those offering critical perspectives on specific issues where ethics was only mentioned in passing.

    \item  Papers on "AI ethics" appeared in large numbers in our results. In our formulation of design practice and keeping our focus on "design ethics", we omitted these papers if they only focus on engineering design, system design, and information design. We included AI ethics and computer ethics papers only if they explicitly involved designers as research participants. 

    \item Because educational contexts involve completely different factors and dynamics and do not fit with our main focus on design practice, we excluded papers focused on teaching ethics to design students, or curriculum studies.

\end{itemize}

\begin{figure} [h!]
    \centering
    \includegraphics[width=0.6\linewidth]{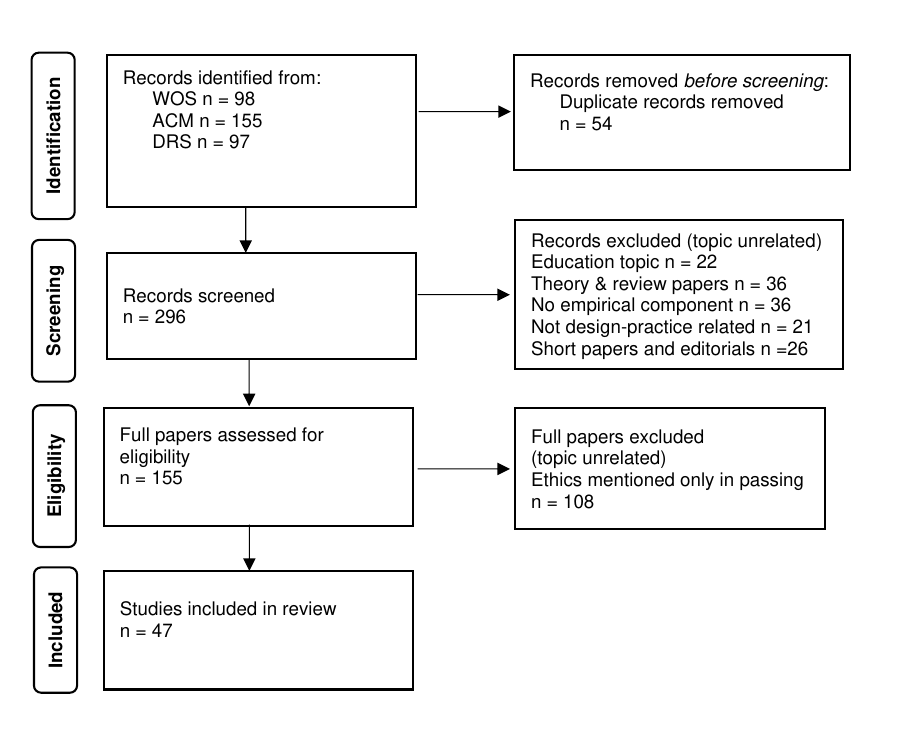}
    \caption{PRISMA flow diagram}
    \label{fig:prisma}
\end{figure}

After applying these criteria, we revisited our publication pool to ensure consistency and alignment with the study’s objectives. All decisions were cross-checked by the authors to maintain accuracy and rigor. Ultimately, 47 papers were selected for detailed review (see Figure \ref{fig:prisma}).

\vspace{5mm} 

\subsection{Data Charting and Synthesizing of Results}
We adopted a descriptive-analytical method \cite{arksey_scoping_2005} to chart the selected studies. This approach involved applying a consistent framework to examine all primary research reports and extracting key details from each study. As a team, we agreed on a set of data items to extract from each paper, ensuring alignment with our main research questions. These items included both general study information and specific details about methods, themes, and findings \cite{arksey_scoping_2005}. We classified each included citation according to: (1) publication year; (2) field of design; (3) context of the research; (4) location of the research; (5) methods used; (6) research conducted by whom, with whom, and for whom; (7) main purpose of the research; (8) approaches or theories of ethics applied; (9) adjacent terms used interchangeably with ethics; (10) main study motivations; (11) main findings; and (12) the type of contribution.

Each author recorded the agreed-upon data items in a shared spreadsheet for all selected studies. When uncertainties arose —such as when a data item did not fit well within the predefined categories or when a new category seemed necessary— we met to discuss and refine our approach. To ensure consistency and accuracy, we also cross-checked each other’s work by randomly reviewing data extracted by other team members.   

To synthesize and summarize the charted data, we identified key themes from the literature based on our research questions and grouped them under thematic headings. For example, data items such as approaches, theories, and adjacent terms were combined under the "Conceptualizations" thematic heading as they are complimentary to each other, while purposes, findings, and contributions were summarized together under the "Purposes and Contributions" thematic heading to highlight their relationships. We used tables to show the distribution of papers by publication year and venue, as well as to present methods, study types, main research purposes, and contributions.

\subsection{Limitations}
While we have aimed to make this review as comprehensive as possible, we acknowledge that some relevant articles may have been missed due to our keyword selection and inclusion criteria. As this is a scoping review, our primary focus has been on maintaining consistency within the dataset rather than including every possible study. Significant efforts have been put into developing and applying inclusion criteria to balance feasibility with comprehensiveness \cite{pare_synthesizing_2015}, as our goal is to have a manageable number of studies that provide a meaningful foundation for understanding the body of research on design ethics in practice.

As the review focuses on ethics in design practice, we have tackled the question of what constitutes "design practice’ throughout the review process. We rely on the authors’ self-positioning of their research; studies are included in the review pool if the authors explicitly state that their research focuses on practice. This reliance may have led to the exclusion of research that, while relevant, did not explicitly frame itself within the context of design practice. Nevertheless, the studies we selected span a variety of practices, and we believe we have captured a level of diversity that enables us to characterize the turn to practice in design ethics.

The focus of the review on empirical research reflects the goal of capturing insights from design practices. This framing led to the exclusion of theoretical and review papers from the final selection, even when these papers were closely related to our topic. We have nonetheless tried to include these papers in the Introduction section to acknowledge their contributions and integrate their insights. Additionally, the search queries retrieved theory-developing papers that exemplify a set of empirical cases, such as in-action ethics \cite{frauenberger_-action_2016}, and we included these in our selection. 

Finally, the review focuses on papers published in the last ten years. Both benchmarked reviews (i.e. \cite{nunes_vilaza_scoping_2022}) and our own search results (see Table \ref{tab:publication_venues}) reveal a noticeable increase in publications, especially after 2020. Thus, we consider this ten-year timeframe to be sufficient to capture the growing emphasis on the turn to practice in design ethics research.

\section{Results}
Before delving into a detailed analysis, we present an overview of key results to contextualize our review. Table \ref{tab:publication_venues} summarizes the distribution of publications across years and venues, highlighting a notable increase in the last five years. During the initial five years (2014–2019), nine papers were identified in our review pool. This number significantly increased in the subsequent five years (2019–2024), reaching 38 publications.

\begin{table}[!ht]
    \centering
    \renewcommand{\arraystretch}{1.2} 
    \setlength{\tabcolsep}{10pt} 
    \begin{tabular}{|l |c| p{8cm}|} 
        \hline
        \textbf{Year} & \textbf{Instances} & \textbf{Venue and References} \\ 
        \hline
        2016 & 2  & CHI \cite{mcnally_childrens_2016}, SIGCAS \cite{gram-hansen_participatory_2016} \\ 
        \hline
        2017 & 5  & Interacting with Computers \cite{malinverni_autoethnographic_2016, shilton_blended_2017, frauenberger_-action_2016}, 
        CHI \cite{barry_mhealth_2017}, ACM SIGDOC \cite{petersen_empathetic_2017} \\ 
        \hline
        2019 & 2  & CHI \cite{gray_ethical_2019}, Co-design \cite{kelly_towards_2019} \\ 
        \hline
        2020 & 6 & Co-design \cite{spiel_details_2020}, Design and Culture \cite{ahmed_minority_2020}, CHI \cite{chivukula_dimensions_2020, madaio_co-designing_2020}, 
        Urban Planning \cite{roosen_dialectical_2020}, NordiCHI \cite{lindberg_cultivating_2020} \\ 
        \hline
        2021 & 7 & CHI \cite{chivukula_identity_2021, wong_timelines_2021}, Design Studies \cite{rivard_articulating_2021}, International Journal of Technoethics \cite{jacobs_bridging_2021}, 
        C\&T \cite{braybrooke_care-full_2021}, CSCW \cite{lindberg_design_2021} PIVOT \cite{lujan_escalante_dancing_2021} \\ 
        \hline
        2022 & 11 & Design Studies \cite{gray_using_2022}, Games and Culture \cite{karlsen_balancing_2022}, International Journal of Design \cite{dindler_engagements_2022, nelissen_rationalizing_2022}, International Journal of Technoethics \cite{lecomte_improving_2022}, CHI \cite{popova_vulnerability_2022, luria_letters_2022}, NordiCHI \cite{pillai_exploring_2022}, Nordes \cite{lindberg_cultivating_2022}, DRS \cite{lujan_escalante_ethics_2022, ozkaramanli_design_2022} \\ 
        \hline
        2023 & 3 & DIS \cite{haghighi_workshop-based_2023, gray_scaffolding_2023}, MobileHCI \cite{yoon_ethical_2023} \\ 
        \hline
        2024 & 11 & Design and Culture \cite{beattie_its_2024}, Visual Communication \cite{cooper_design_2024}, CHI \cite{gray_building_2024, mildner_listening_2024}, NordiCHI \cite{popova_voicing_2024}, ACM Games \cite{millard_ethics_2024}, ACM J. Responsib. Comput. \cite{gray_languaging_2024}, Proc. ACM Hum.-Comput. Interact. \cite{popova_who_2024, lindberg_doing_2024}, DRS \cite{chivukula_envisioning_2024, tekogul_cultivating_2024}\\ 
        \hline
    \end{tabular}
    \caption{Summary of publications by year and venue, 47 papers in total}
    \label{tab:publication_venues}
\end{table}

When examining publication venues, 28 papers (60 \%) were published in design in HCI-related venues such as DIS, CHI, and NordiCHI. Furthermore, as detailed in Table \ref{tab:fields}, 26 papers (55\%) specifically position themselves within design practices in HCI, focusing on design practices such as UX design and GUI design. Many of these studies employ broad terms such as "technology design" or "technology and design practitioners" to frame their work, often emphasizing digital and interactive design practices. The participatory design (PD) field is also prominent. Some studies focus on ethics in PD more generally \cite{kelly_towards_2019, gram-hansen_participatory_2016}, while others examine the ethical considerations of designing with specific user groups, such as teenagers and children \cite{mcnally_childrens_2016, spiel_details_2020, malinverni_autoethnographic_2016}. Contributions from other domains, such as industrial design and graphic design, are relatively infrequent. While research in the "Various" group involves participants from multiple design fields, research in the "Not Specified" group does not mention any specific design field and refers to design in general terms. When combining insights from venues and design fields, it is evident that design in HCI-related research significantly contributes to the emerging discourse on design ethics in practice. Our analysis will further explore the motivations behind this engagement, both in terms of the increasing number of studies and field-specific focus, as well as its potential implications.

\begin{table} [!ht]
    \centering
    \renewcommand{\arraystretch}{1.2} 
    \setlength{\tabcolsep}{10pt} 
    \begin{tabular}{|l|c|p{6cm}|}
    \hline
        \textbf{Design Fields} & \textbf{Instances} & \textbf{Design Practices and References}\\
        \hline
        Design practices within HCI & 26 & Soma Design \cite{popova_voicing_2024,popova_vulnerability_2022}, UX Design \cite{yoon_ethical_2023, lindberg_cultivating_2020, lindberg_design_2021, chivukula_dimensions_2020, gray_ethical_2019, pillai_exploring_2022, beattie_its_2024, nelissen_rationalizing_2022, wong_timelines_2021}, Technology Design \cite{gray_scaffolding_2023, gray_building_2024, haghighi_workshop-based_2023, chivukula_identity_2021, chivukula_envisioning_2024, gray_languaging_2024, dindler_engagements_2022, shilton_blended_2017, frauenberger_-action_2016, barry_mhealth_2017, madaio_co-designing_2020, popova_who_2024}, Game Design \cite{karlsen_balancing_2022, millard_ethics_2024}, Conversational User Interface \cite{mildner_listening_2024} \\
        \hline
        Participatory Design & 5 & \cite{mcnally_childrens_2016, spiel_details_2020, kelly_towards_2019, gram-hansen_participatory_2016, malinverni_autoethnographic_2016}\\
        \hline
        Other & 12 & Industrial Design \cite{lecomte_improving_2022}, Graphic Design \cite{ahmed_minority_2020}, Communication Design \cite{cooper_design_2024, petersen_empathetic_2017}, Urban Design \cite{roosen_dialectical_2020}, Method Design \cite{gray_using_2022}, Various \cite{braybrooke_care-full_2021, jacobs_bridging_2021, rivard_articulating_2021, tekogul_cultivating_2024, lindberg_doing_2024, lindberg_cultivating_2022} \\
        \hline
        Not Specified & 4 & \cite{lujan_escalante_dancing_2021, ozkaramanli_design_2022, luria_letters_2022, lujan_escalante_ethics_2022} \\
        \hline
    \end{tabular}
    \caption{Summary of design practices in the focus of the research stated by the authors}
    \label{tab:fields}
\end{table}

Finally, while we aimed to categorize research by location, this proved challenging as most studies do not explicitly indicate the socio-geographical context of their research. It was often unclear whether the research context aligned with the location of the authors and their institutions, the origins of the study participants, or the contexts of the companies involved. Consequently, it can be said that there is a noticeable lack of contextualization regarding the geographical and cultural settings of the research, as well as the researchers' positionality within those settings.

\section{Analysis}

In this section, we analyze the review data in response to our research questions, focusing on the motivations behind the studies, the theoretical approaches and empirical methods employed, as well as their primary research purposes and contributions. While presenting our results, we also identify the common characteristics of how the concept of design ethics is defined in these studies, highlight existing gaps, and explore the potentials emerging from the review.

\subsection{Study Motivations}
To identify the motivations for studying design ethics in practice, we will examine researchers’ perspectives on why they engage with this area, why they consider it important and timely, and what factors may be driving the increasing number of studies published each year. After considering the motivations for studying design ethics, we will explore why researchers are turning to practice in this field.

Along with acknowledging the long-established body of research examining how design methods and objects embody specific beliefs, values, and discourses \cite{haghighi_workshop-based_2023, malinverni_autoethnographic_2016, lindberg_cultivating_2020, lindberg_cultivating_2022}, several studies - particularly those from the HCI research - in our review pool highlight the growing ethical responsibility of design in the context of emerging technologies and their widespread impact on daily life \cite{madaio_co-designing_2020, pillai_exploring_2022, lecomte_improving_2022, luria_letters_2022, chivukula_identity_2021, wong_timelines_2021, frauenberger_-action_2016, dindler_engagements_2022, haghighi_workshop-based_2023, ozkaramanli_design_2022, lindberg_cultivating_2020, cooper_design_2024, ahmed_minority_2020}. These sources express concerns about potential negative consequences of technologies, such as algorithmic bias, surveillance, privacy issues, and the amplification of societal biases; they emphasize the need for proactive ethical considerations throughout the design process to anticipate potential negative consequences, and to develop ethical frameworks that can be utilized in design practice to support this. As an example, Haggighi et al. \cite{haghighi_workshop-based_2023} illustrate how advancements in ubiquitous computing have influenced design ethics by developing a workshop-based method to explore ethical implications. Due to new questions raised by technological change—such as the collection and use of personal data—researchers were motivated to develop a new method for addressing emerging ethical challenges \cite{haghighi_workshop-based_2023}. Additionally, amid the growing ethical responsibility prompted by emerging technologies, particular attention is given to the responsibility toward users, emphasizing how the design of technology directly impacts users’ behaviors and experiences \cite{mildner_listening_2024, petersen_empathetic_2017}. This body of research promotes user-centered approaches \cite{rivard_articulating_2021}, scrutinizes "manipulative" UX design tactics \cite{beattie_its_2024}, and identifies challenges faced by designers advocating for users while balancing commercial demands \cite{nelissen_rationalizing_2022, beattie_its_2024}.

On a different note, when it comes to design ethics in participatory design (PD), some researchers reason their work based on the inherent qualities of PD itself \cite{mcnally_childrens_2016, spiel_details_2020, kelly_towards_2019, gram-hansen_participatory_2016, malinverni_autoethnographic_2016}. In this body of work, there is no specific emphasis on changes brought about by technological advancements. Instead, the collaborative and empowerment-driven nature of PD is seen as inherently raising ethical considerations—such as managing diverse perspectives, ensuring equitable power distribution, and addressing the social impact of design decisions.

Nonetheless, for the majority, research on design ethics in practice is driven by a sense of urgency due to rapidly growing digital technologies to ensure that design’s increasing power and influence are wielded ethically and responsibly. This motivation reflects into a heightened focus on design ethics in practice to be able to produce research that can be translated into and utilized in everyday design work to uncover, examine, and embed values into design artifacts and technical elements \cite{chivukula_dimensions_2020, lindberg_design_2021, lindberg_cultivating_2022, jacobs_bridging_2021, barry_mhealth_2017, spiel_details_2020}.  

There is a well-recognized gap between theoretical discussions of design ethics and the practical experiences of designers \cite{lindberg_design_2021, lindberg_cultivating_2022, jacobs_bridging_2021}. Academic research often struggles to translate into readily applicable tools and approaches for practitioners. Studying ethics in practice helps bridge this gap by grounding research in real-world contexts, generating insights that are relevant and actionable for designers. Sources employ ethics in practice to "resonate" \cite{gray_scaffolding_2023}, "align" \cite{lindberg_design_2021}, and "attune" \cite{dindler_engagements_2022} with the everyday realities of design practitioners. This approach also helps to "translate identified values into technical affordances" \cite{shilton_blended_2017} and create "actionable interventions" \cite{wong_timelines_2021}.

\subsection{Ethical Conceptualizations}
In this section, we explore the prevailing theories and concepts in the literature and the ways in which they are applied. Of the articles reviewed, 24 (51 \%) employ theories and concepts that explicitly ground a practice-oriented approach. Several terms are used to describe the conceptual frameworks, including "in-action ethics" \cite{frauenberger_-action_2016}, "ethics on the ground" \cite{gray_ethical_2019}, "micro-ethics" \cite{spiel_details_2020}, "everyday ethics" \cite{gray_ethical_2019, lindberg_design_2021}, "situated ethics" \cite{spiel_details_2020, malinverni_autoethnographic_2016}, "ethics in real-time" \cite{roosen_dialectical_2020}, "pragmatist ethics" \cite{gray_languaging_2024, gray_scaffolding_2023, lindberg_cultivating_2020}, and "processual ethics" \cite{popova_vulnerability_2022}. Though these concepts differ in meaning, they may fall under the broader turn to practice in design ethics research, sharing key traits in their approach to design ethics. Four main attributes of design ethics emerge from a thematic grouping of how these frameworks interpret the concept of ethics: 

\textit{Ethics are inherent to design practices:} Research sees ethics as something separate that is added onto design work, but rather as an integral and inseparable part of the design process itself.  Design choices inherently carry ethical weight, shaping the values and norms embedded in technologies and affecting how people interact with the world. This recognizes that design is a form of "doing ethics" \cite{dindler_engagements_2022, frauenberger_-action_2016, popova_voicing_2024}.

\textit{Ethics are contextually dependent:} The sources strongly emphasize the situated nature of design ethics. This means that ethical considerations in design are not absolute or universal; rather, they are shaped by the specific context in which design occurs \cite{frauenberger_-action_2016, haghighi_workshop-based_2023, rivard_articulating_2021}. This is a dynamic and context-sensitive process that is shaped by various factors, including the designer’s values, organizational culture, stakeholder needs, and the specifics of the design project \cite{dindler_engagements_2022, lindberg_design_2021, popova_who_2024}.

\textit{Ethics are not static but evolving:} The review highlights that ethical considerations are dynamic, evolving over time. This evolving nature of ethics in design practice challenges rigid, rule-based approaches and points to the need to adopt a more responsive, flexible understanding, as well as underlines the temporality aspect of ethical encounters \cite{frauenberger_-action_2016, lindberg_cultivating_2020}

\textit{Ethics can be cultivated:} Several sources argue that ethical practices can be developed and nurtured over time. Ethical practices, therefore, can be negotiated and co-constructed within the context of each project, as well as within wider design cultures \cite{ozkaramanli_design_2022, lindberg_design_2021, lujan_escalante_ethics_2022}. 

Overall, the design ethics in practice corpus emphasizes the situated and fluid nature of design ethics and thus the importance of moving beyond abstract principles and theoretical frameworks to observe how ethics are enacted in actual design settings, calling for practice-oriented research. For example, Frauenberger et al. \cite{frauenberger_-action_2016}, through a case on privacy issues regarding children's data,  illustrate how design ethics require real-time decisions and constant reflection. Their paper demonstrates that the initial ethical plan could not address the complexities of family dynamics, highlighting the limits of pre-defined and rule-centered approaches to handling design ethics. Their approach involves understanding the complexities, contradictions, and compromises that arise when designers grapple with ethical considerations in their daily work.

The review highlights some common notions and concepts that researchers use to position their work. The first concept we can highlight is the third paradigm HCI mentioned several times as an approach that has shaped researchers’ perspectives on ethics research. As mentioned, the third paradigm HCI is marked by a shift toward situated perspectives that acknowledge the complex, dynamic, and value-laden nature of HCI \cite{malinverni_autoethnographic_2016, frauenberger_-action_2016}. Some sources \cite{frauenberger_-action_2016, dindler_engagements_2022, malinverni_autoethnographic_2016, lindberg_cultivating_2020, gray_ethical_2019} explicitly suggest that this shift has influenced their design ethics research, promoting a more situated, dynamic, and practice-oriented approach. This shift encourages researchers to explore the complexities of ethical decision-making within real-world design contexts, collaborating with practitioners to develop methods and frameworks that support the cultivation of ethical awareness, responsibility, and responsiveness within design teams and organizations.

Another concept frequently encountered in design ethics research within the HCI field is "dark patterns" \cite{beattie_its_2024, mildner_listening_2024, nelissen_rationalizing_2022, gray_ethical_2019, pillai_exploring_2022, millard_ethics_2024}. This concept refers to user-interface design choices that manipulate or deceive users into making decisions they would not make if they were fully informed and aware of alternatives. It is stated that dark patterns exploit cognitive biases and well-established design principles to subtly influence user behavior, often for commercial gain. They are identified by a lack of transparency, trust, and respect for user autonomy, ultimately disregarding ethical considerations in design. Some sources \cite{beattie_its_2024, mildner_listening_2024, nelissen_rationalizing_2022} suggest that the pervasiveness of dark patterns underscores the need for greater ethical awareness and responsibility within the design community.

The next concept repeatedly encountered in the literature is the "ethics of care". Studies merge situated ethics and the ethics of care \cite{rivard_articulating_2021, braybrooke_care-full_2021, tekogul_cultivating_2024, ozkaramanli_design_2022, petersen_empathetic_2017, popova_voicing_2024, popova_vulnerability_2022, spiel_details_2020, popova_who_2024}, framing ethics as an "invitation to care". This approach involves fostering a sense of empathy toward collaborators and interpreting emerging ethical issues by attending to the needs and interests of others involved in the design process. This line of work also highlights the role of the body and emotions in relation to ethics. Certain sources \cite{popova_voicing_2024, popova_vulnerability_2022, popova_who_2024, lujan_escalante_dancing_2021} point to a move away from purely rational and theoretical frameworks toward a deeper recognition of the role of the body and emotions in shaping ethical understanding and practice.

Other concepts encountered in the literature include "constructive ethics" \cite{gram-hansen_participatory_2016}, which draws upon Løgstrup’s philosophy to emphasize the ethical demands inherent in design, and "phronesis" \cite{barry_mhealth_2017}, a concept from Aristotelian virtue ethics referring to practical wisdom or judgment acquired through experience in specific contexts. Technology-specific concepts also appear, such as "AI ethics" \cite{yoon_ethical_2023}, "computer ethics" \cite{shilton_blended_2017}, "mixed reality ethics" \cite{millard_ethics_2024}.

Finally, it is important to note that many studies do not define the concept of ethics or explicitly state what is considered ethical. Instead, they rely on approaches that guide their research practice and perspective on design ethics. This is not surprising, given that our review focuses on research with an empirical component—prioritizing insights from practice over pre-established definitions. A few studies  \cite{karlsen_balancing_2022, lindberg_cultivating_2020, chivukula_dimensions_2020, gray_ethical_2019} explicitly state that they avoid defining ethics because they aim to understand how designers themselves define it.

\subsection{Methods Used}
This section will examine different methods employed in the literature and the methods-related findings of our review. Because conducting empirical research was a key filtering criterion for this review, the methods observed align with this focus. While categorizing the methods used, we will also examine how the research utilizes these methods to address the hands-on, everyday quality of ethics and how they capture the implicit and often tacit aspects of ethical reasoning that may arise in the design process. In addition to these observational methods, we will also consider research that uses interventional methods aimed at cultivating and operationalizing ethics within the design process.

Table \ref{tab:method} shows that the interview method is the most commonly used approach in the reviewed studies. If we include mixed-methods studies with an interview component, around 51\% of the research incorporates interviews as a data-collection method. Researchers use interviews to gain insights into practitioners’ perspectives on ethics in their work and organizations, to identify ethical challenges and opportunities within specific design contexts \cite{dindler_engagements_2022, rivard_articulating_2021, lindberg_design_2021, jacobs_bridging_2021, chivukula_dimensions_2020, beattie_its_2024, millard_ethics_2024}, and to explore specific dimensions of ethical decision-making, such as the affective dimension \cite{popova_who_2024}, identity dimension \cite{chivukula_identity_2021}, and business models \cite{karlsen_balancing_2022}. Among the interview studies, some incorporate designerly approaches into the interview process, such as speculative enactments \cite{nelissen_rationalizing_2022}, probes studies \cite{luria_letters_2022, popova_vulnerability_2022}, and visual trigger materials \cite{lindberg_design_2021} to stimulate conversation, address difficult topics, and encourage future-oriented thinking during interviews.

\begin{table} [h!]
    \centering
    \renewcommand{\arraystretch}{1.2} 
    \setlength{\tabcolsep}{10pt} 
    \begin{tabular}{|l|c|p{6cm}|}
    \hline
        \textbf{Method} & \textbf{Instances} & \textbf{References}\\
        \hline
        Interviews & 16 & \cite{rivard_articulating_2021, karlsen_balancing_2022, jacobs_bridging_2021, lindberg_design_2021, chivukula_dimensions_2020, petersen_empathetic_2017, dindler_engagements_2022, chivukula_identity_2021, mildner_listening_2024, kelly_towards_2019, gray_using_2022, popova_who_2024, beattie_its_2024, nelissen_rationalizing_2022, millard_ethics_2024, gray_scaffolding_2023} \\
        \hline
        Workshops & 11 & \cite{lindberg_cultivating_2020, lindberg_cultivating_2022, lindberg_doing_2024, popova_voicing_2024, chivukula_envisioning_2024, gray_building_2024, roosen_dialectical_2020, gram-hansen_participatory_2016, pillai_exploring_2022, gray_languaging_2024, haghighi_workshop-based_2023} \\
        \hline
        Case Studies & 6 & \cite{lecomte_improving_2022, barry_mhealth_2017, shilton_blended_2017, spiel_details_2020, lujan_escalante_ethics_2022, frauenberger_-action_2016} \\
        \hline
        Mixed-Methods & 8 & Interview + Observations \cite{gray_ethical_2019, tekogul_cultivating_2024, yoon_ethical_2023} \newline Interview + Survey \cite{ahmed_minority_2020, mcnally_childrens_2016} \newline Interview + Probes Study \cite{luria_letters_2022, popova_vulnerability_2022} \newline Interview + Workshop \cite{madaio_co-designing_2020} \\
        \hline
        Other & 6 & Autoethography \cite{malinverni_autoethnographic_2016}, Conversation \cite{ozkaramanli_design_2022}, Dancing \cite{lujan_escalante_dancing_2021}, Design Sprint \cite{braybrooke_care-full_2021}, Design Activity \cite{wong_timelines_2021, cooper_design_2024} \\
        \hline
    \end{tabular}
    \caption{Summary of methods employed}
    \label{tab:method}
\end{table}

There are six case studies within the review pool. Three research papers present multiple case studies together to offer a broader perspective on the design ethics \cite{frauenberger_-action_2016, spiel_details_2020, shilton_blended_2017}, while single case studies focus on a specific company \cite{lecomte_improving_2022}, or a specific product development process \cite{barry_mhealth_2017, lujan_escalante_ethics_2022}. For example, Shilton and Anderson \cite{shilton_blended_2017} use three ethnographic case studies to illustrate the different ways of enacting the concept of values levers, which are development practices that facilitate conversations about values and build consensus around their importance in design. Their findings highlight that (1) having an ethical design advocate role in a design team helps normalize ethics, (2) understanding a team's existing values and assumptions helps guide meaningful discussions, and (3) open conversations and consensus-building prove more effective than enforcing external rules.

The literature also contains a variety of less common methods that derive from design and arts research, such as designing activities \cite{cooper_design_2024, wong_timelines_2021}, using dancing as a method \cite{lujan_escalante_dancing_2021}, employing design sprints \cite{braybrooke_care-full_2021}, and use of probes and fictional scenarios in interviews \cite{luria_letters_2022, nelissen_rationalizing_2022}. As shown in Table \ref{tab:type},  36\% (17 studies) employed interventional methods such as co-creation workshops and soma design workshops. Co-creation and co-design workshops often involve hands-on activities in which participants collaboratively develop tangible outputs, such as ethical toolkits \cite{lindberg_cultivating_2020}, action plans \cite{gray_building_2024, chivukula_envisioning_2024}, checklists \cite{madaio_co-designing_2020}, and design artifacts \cite{lindberg_cultivating_2022, haghighi_workshop-based_2023} to bring together researchers and practitioners to explore ethical issues in design collaboratively. For example, Gray et al. \cite{gray_building_2024} held co-creation workshops with design and technology practitioners to help them create personalized ethics-related action plans. The workshops aimed to empower participants to tackle ethical challenges using their design expertise. They identified key roles participants took on (i.e., the advocate, the operationalizer, the reformer) and three strategies they used to define their ethical problem space (i.e., refining, expanding, diverging). Another study used body-based workshops to explore the concept of "felt ethics", centering around the idea that ethics can be experienced as a bodily sensation \cite{popova_voicing_2024}. The workshop methodology is grounded in a soma design approach, which uses the body as a site for exploration and reflection on design practice to probe themes like discomfort, boundaries, control, and negotiation.

\begin{table} [h!]
    \centering
    \renewcommand{\arraystretch}{1.2} 
    \setlength{\tabcolsep}{10pt} 
    \begin{tabular}{|c|c|}
    \hline
        \textbf{Interventional} & \textbf{Observational} \\
        \hline
        36\% & 64\% \\
        \hline
    \end{tabular}
    \caption{Distribution of method types}
    \label{tab:type}
\end{table}

Finally, we observe a greater prevalence of cross-sectional methods compared to longitudinal ones (see Table \ref{tab:time}). Cross-sectional research involves collecting data from each participant at a single point in time, providing a snapshot of a situation—for example, through interviews. Repeated cross-sectional research collects data at multiple time points, typically from different samples, or occasionally from the same participants in different settings over short durations, such as in a series of workshops. Longitudinal research involves collecting data from the same participants across multiple time points over an extended period, with the goal of studying changes and processes as they unfold over time—for instance, through participant observation \cite{karapanos_advances_2021}. Current research mainly relies on cross-sectional methods, such as interviews, and repeated cross-sectional methods, such as conducting a series of workshops. However, we note the limited number of longitudinal studies that capture the temporal aspects of ethics. We see this as a significant gap in the research, especially given the emphasis on the situated and evolving nature of design ethics.

\begin{table} [h!]
    \centering
    \begin{tabular}{|c|c|c|}
    \hline
        \textbf{Cross-sectional} & \textbf{Repeated cross-sectional} & \textbf{Longitudinal} \\
        \hline
        62\% & 30\% & 8\%\\
        \hline
    \end{tabular}
    \caption{Distribution of study designs}
    \label{tab:time}
\end{table}

\subsection{Purposes and Contributions }

In this section, we will identify the primary research purposes, draw parallels between these purposes and the contributions of the studies, and provide examples of the research findings. We coded the main purposes into four nonexclusive thematic categories, recognizing that a single study can fall under multiple categories simultaneously, which are: understand (how design ethics unfolds in everyday practice), support (practitioners in addressing ethical situations), operationalize (ethical considerations within design practice and organizations), and cultivate (ethical foundations and cultures of design ethics). Table \ref{tab:purpose} shows the distribution of research numbers according to their primary purposes; highlighting the emerging nature of the field, as the majority of research is directed towards understanding the dynamics of ethics in practice.

\begin{table} [h!]
    \centering
     \renewcommand{\arraystretch}{1.2} 
    \setlength{\tabcolsep}{10pt} 
    \begin{tabular}{|p{5cm}|c|p{5cm}|}
    \hline
        \textbf{Purpose} & \textbf{Instances} & \textbf{References}\\
        \hline
        Understand (how design ethics unfolds in everyday practice) & 29 & \cite {ahmed_minority_2020, petersen_empathetic_2017, gray_using_2022, rivard_articulating_2021, karlsen_balancing_2022, dindler_engagements_2022, mcnally_childrens_2016, kelly_towards_2019, lindberg_design_2021, beattie_its_2024, chivukula_identity_2021, yoon_ethical_2023, popova_who_2024, ozkaramanli_design_2022, millard_ethics_2024, gray_languaging_2024, popova_vulnerability_2022, lindberg_doing_2024, chivukula_envisioning_2024, gray_ethical_2019, gray_building_2024, madaio_co-designing_2020, nelissen_rationalizing_2022, shilton_blended_2017, mildner_listening_2024, gray_scaffolding_2023, chivukula_dimensions_2020, pillai_exploring_2022, luria_letters_2022}\\
        \hline
        Support (practitioners in addressing \newline ethical situations) & 17 & \cite{lujan_escalante_dancing_2021, popova_voicing_2024, roosen_dialectical_2020, malinverni_autoethnographic_2016, wong_timelines_2021, braybrooke_care-full_2021, cooper_design_2024, lecomte_improving_2022, haghighi_workshop-based_2023, spiel_details_2020, nelissen_rationalizing_2022, shilton_blended_2017, mildner_listening_2024, gray_scaffolding_2023, chivukula_dimensions_2020, pillai_exploring_2022, luria_letters_2022} \\
        \hline
        Operationalize (ethical considerations within practice and organizations)  & 10 & \cite{lindberg_cultivating_2022, lujan_escalante_ethics_2022, jacobs_bridging_2021, gram-hansen_participatory_2016, barry_mhealth_2017,frauenberger_-action_2016, spiel_details_2020, chivukula_envisioning_2024, gray_ethical_2019, madaio_co-designing_2020} \\
        \hline
        Cultivate (ethical foundations and \newline cultures of design ethics) & 3 & \cite{tekogul_cultivating_2024, lindberg_cultivating_2020, lindberg_cultivating_2022}\\
        \hline
    \end{tabular}
    \caption{Categories of main research purposes}
    \label{tab:purpose}
\end{table}

For categorizing contributions, we follow Wobbrocks’ article \cite{wobbrock_research_2016}, which identifies seven types of contribution in HCI research: empirical, methodological, survey, artifact, dataset, theoretical, and opinion. Given that one of our primary inclusion criteria is that studies must include an empirical component, all the works reviewed provide empirical contributions. Additionally, we encountered (i) methodological contributions, including the development of methods to support ethical practice, (ii) theoretical contributions derived from empirical insights (See Table \ref{tab:contributions}), and (iii) one artifact contribution where a website designed to assist practitioners in finding relevant ethical decision-making tools tailored to their specific contexts \cite{gray_scaffolding_2023}. Now, we delve into each purpose category in detail, aligning them with their corresponding contribution categories and key findings. This approach ensures a clear connection between the purpose, its contributions, and the insights gained.

\begin{table} [h!]
    \centering
    \renewcommand{\arraystretch}{1.2} 
    \setlength{\tabcolsep}{10pt} 
    \begin{tabular}{|l|c|p{6cm}|}
    \hline
        \textbf{Contribution} & \textbf{Instances} & \textbf{References}\\
        \hline
        Empirical & 26 & \cite{tekogul_cultivating_2024, rivard_articulating_2021, karlsen_balancing_2022, shilton_blended_2017, gray_building_2024, mcnally_childrens_2016, madaio_co-designing_2020, lindberg_cultivating_2022, ozkaramanli_design_2022, chivukula_dimensions_2020, petersen_empathetic_2017, dindler_engagements_2022, chivukula_envisioning_2024, yoon_ethical_2023, pillai_exploring_2022, chivukula_identity_2021, beattie_its_2024, ahmed_minority_2020, millard_ethics_2024, kelly_towards_2019, gray_using_2022, popova_vulnerability_2022, popova_who_2024, lindberg_doing_2024, lindberg_design_2021, gray_languaging_2024} \\
        \hline
        Empirical + Methodological & 11 & \cite{haghighi_workshop-based_2023, malinverni_autoethnographic_2016, braybrooke_care-full_2021, lindberg_cultivating_2020, lujan_escalante_dancing_2021, cooper_design_2024, lecomte_improving_2022, luria_letters_2022, nelissen_rationalizing_2022, wong_timelines_2021, popova_voicing_2024} \\
        \hline
        Empirical + Theoretical & 9 & \cite{gram-hansen_participatory_2016, mildner_listening_2024, jacobs_bridging_2021, gray_ethical_2019, lujan_escalante_ethics_2022, frauenberger_-action_2016,barry_mhealth_2017, spiel_details_2020, roosen_dialectical_2020} \\
         \hline
        Empirical + Artifact & 1 & \cite{gray_scaffolding_2023} \\
        \hline
    \end{tabular}
    \caption{Types of main research contributions}
    \label{tab:contributions}
\end{table}

\subsubsection{Understand}
This category contains research that aims to understand how designers engage with ethics in their practice and how different factors affect ethical positioning within practice. The contribution type in this form of inquiry is mainly empirical. 

These studies aim to gather insights into how designers perceive and engage with ethics in their work, as well as the concepts, values, and challenges shaping their ethical decision-making process \cite{lindberg_design_2021, dindler_engagements_2022, chivukula_envisioning_2024, chivukula_dimensions_2020}. By examining design practices, these studies identify factors influencing ethical outcomes, such as the role of identity and belief in ethical positioning \cite{chivukula_identity_2021}, the affective dimensions of ethics \cite{popova_who_2024, popova_vulnerability_2022}, team dynamics around ethics \cite{shilton_blended_2017}, and the vocabularies practitioners use to negotiate ethical concepts \cite{pillai_exploring_2022, gray_languaging_2024}. Additionally, they explore the ethics of participation \cite{kelly_towards_2019, mcnally_childrens_2016, spiel_details_2020}, as well as ethics in specific technologies, including artificial intelligence  \cite{yoon_ethical_2023, madaio_co-designing_2020}, mixed reality \cite{millard_ethics_2024}, and conversational interfaces \cite{mildner_listening_2024} and issues such as privacy \cite{beattie_its_2024, nelissen_rationalizing_2022}, and monetization \cite{karlsen_balancing_2022}. 

One of the prominent findings to highlight in this area is that practitioners often perceive ethics as a vague and challenging concept to define. As previously noted, some researchers choose not to define ethics in their studies, aiming instead to explore how designers themselves define it; however, many practitioners struggle to articulate individual definitions, often relying on adjacent terms like sustainability or privacy to describe ethical considerations \cite{beattie_its_2024, dindler_engagements_2022}. Beattie et al. \cite{beattie_its_2024} describe ethics as a "floating signifier", a term lacking fixed meaning and easily attached to various domains or concepts, illustrating the inherent ambiguity of the concept in practice. 

\subsubsection{Support}
Research in this category aims to support practitioners in addressing ethical dilemmas, implications, and decisions in their work by making ethical considerations more accessible and actionable in everyday design practice. The contributions in this area are primarily methodological and empirical. 

This type of research can develop methods such as visual mapping methods \cite{cooper_design_2024, wong_timelines_2021, roosen_dialectical_2020}, body-based methods \cite{lujan_escalante_ethics_2022, popova_voicing_2024}, or speculative design methods \cite{haghighi_workshop-based_2023, luria_letters_2022, nelissen_rationalizing_2022}. They can build collaborative and codesign spaces \cite{braybrooke_care-full_2021, lindberg_cultivating_2022} to tackle the complex and context-dependent nature of ethics in design practice. For example, "design timescapes" \cite{cooper_design_2024} is a method to help designers visualize the past, present, and future of their work, linking it to societal changes for better-informed decisions. Another example can be "timelines" \cite{wong_timelines_2021}; a method that uses world-building to explore the ethical implications of technologies or artifacts through narratives. Combining design fiction, scenario planning, and value-sensitive design, it claims to offer a flexible way to advocate for values and examine ethics.

These studies share a common finding: while a wide range of tools and methods exist to support designers in addressing ethical concerns, they are often underused in practice. Many ethical design tools and methods are perceived as overly abstract or disconnected from the realities of daily design work. Practitioners often find that academic frameworks for ethical design fail to translate into actionable steps that can be integrated into their workflows \cite{gray_building_2024, lindberg_cultivating_2020, lindberg_design_2021, rivard_articulating_2021, madaio_co-designing_2020}. Research in this category specifies that support mechanisms with insights from actual lived experiences should be built to bridge this gap. For example, a prominent term, "practice resonance"  \cite{gray_scaffolding_2023}, refers to the degree to which a design method aligns with the practical needs, values, and work contexts of practitioners, thereby enabling them to effectively engage with and apply the method in their everyday work. This term underscores the importance of grounding ethics research in practical contexts. The authors argue that methods should be designed to support practitioners in "successfully navigating, perturbing, and potentially resolving aspects of their ethical complexity" \cite{gray_building_2024}, which is shaped by organizational contexts, disciplinary roles, and existing ethical knowledge.

\subsubsection{Operationalize}
Research in this category positions ethics as a form of mediation within the design process and focuses on how to operationalize ethical considerations within design work and organizational structures. Unlike research in the Support category, which develops tools and methods to help practitioners address ethical dilemmas, studies in the Operationalize category aim to render ethics into actions. This body of work aims to bridge the gap between abstract ethical principles and their practical implementation in design, ensuring that ethical considerations are consciously utilized at every stage of the design process \cite{gram-hansen_participatory_2016, barry_mhealth_2017, gray_building_2024, chivukula_envisioning_2024, gray_ethical_2019, spiel_details_2020}. While support tools are often intended for individual or team use, research in the "Operationalize" category also aims to resonate within and across organizations. The main contributions of this type of research are theoretical and empirical.

As an example, the concepts of ethical mediation and ethical design complexity developed by Gray, Chivukula, and their collaborators are prominent contributions in which they "position ethics as an important mediator of design complexity" \cite{gray_ethical_2019}. Through a series of empirical studies, they construct the concept of ethical design complexity, defined as "complex and choreographed arrangements of ethical considerations" that practitioners must navigate. This complexity emerges from the interplay of individual values and beliefs, organizational practices and cultures, and applied ethical frameworks. In essence, in this context, ethical mediation refers to the ongoing process of navigating this complexity by recognizing multifaceted ethical challenges, engaging in reflective dialogue, and taking actions aligned with responsible and impactful design practices \cite{gray_building_2024, chivukula_dimensions_2020, gray_ethical_2019}. By elaborating on how complexity is structured and situating ethical mediation as the mechanism to navigate it, their work provides a theoretical and practical framework to operationalize design ethics.

Another important example is micro-ethics, as expanded by Spiel et al. \cite{spiel_details_2020} through a series of participatory design processes in which designers collaborate closely with participants, often in sensitive settings. They position ethics as situated and processual, emphasizing care, empathy, and responsibility in interactions. Micro-ethics calls for designers to be more attentive and responsive to the ethical nuances of their interactions and to cultivate an ethical awareness that permeates the entire design process. It operationalizes ethics by making the often unspoken or overlooked ethical aspects of the process more visible, focusing on the small, everyday decisions and actions that unfold throughout the design process.

\subsubsection{Cultivate}
Within the context of design practice, cultivating ethics refers to intentionally fostering an environment and mindset in which ethical considerations are deeply embedded and continuously nurtured throughout the design process. Unlike addressing ethical considerations as they arise, cultivation proactively engages in ethos building. A primary method is conducting interactive workshops, often using a co-design or co-creation approach. These sessions bring practitioners together to explore the meaning of design ethics and related challenges, discuss and co-design possible solutions for dealing with ethical tensions, identify problems and hindrances to ethically responsible design and suggest approaches to overcome them, explore how designers would conduct cultivation of their own ethical practices \cite{lindberg_cultivating_2020, lindberg_cultivating_2022}. This type of research contributes both methodologically and empirically. 

For example, following Frauenberger et al.’s in-action ethics \cite{frauenberger_-action_2016}, a collection of works by Lindberg and colleagues \cite{lindberg_cultivating_2020,lindberg_cultivating_2022, lindberg_design_2021} engages with the concept of ethos building and expands it towards creating an ethical foundation and ethical culture. Ethos building involves a conscious and sustained effort to create a shared understanding of ethical values, integrate those values into the design process, and foster a culture that supports ethical impact. Lindberg et al. emphasize ethos building as (1) a process that involves ongoing reflection, dialogue, and negotiation, (2) a collaborative effort among designers, stakeholders, and the broader community, and (3) a situated practice in which the specific values and commitments that constitute an ethos are shaped by context.

\section{Discussion}
In recent years, there has been a noticeable rise in practice-oriented research on design ethics within HCI and design fields, as reflected in the numbers presented in our review (see Table \ref{tab:publication_venues}). We believe this presents a timely opportunity to consolidate these efforts and reflect on their shared motivations, conceptual foundations, key concerns, and methodological approaches. The studies in our review appear to align in meaningful ways, forming a relatively coherent strand of research. However, despite their prominence, these works remain dispersed across various venues and design disciplines. The turn to practice in HCI provides a useful framing that allows us, through the notion of practice, to characterize these research efforts into a more integrated strand and pinpoint future research directions.

This review synthesizes existing literature on ethics from a practice-oriented perspective across design fields. In addition to ACM DL, we deliberately included both the WOS and DRS libraries, observing traces of ethics in practice in other design disciplines (see Table \ref{tab:publication_venues} and \ref{tab:fields}), in order to encourage cross-pollination between fields for a richer understanding. Nonetheless, it is clear that there are considerably more studies in our pool from HCI research. There may be several reasons for this. (1) As noted, much of the research is motivated by the increasing influence of emerging technologies \cite{pillai_exploring_2022,  chivukula_identity_2021, dindler_engagements_2022, lindberg_cultivating_2020}. Although these technologies impact all areas of work, the focus on emerging digital technologies makes practice-oriented ethics in HCI research more dominant, as designers are often seen as the creators of these technologies. (2) The shift toward practice in HCI studies, as previously discussed, has further supported and shaped design ethics research in this domain. For these reasons, we believe that HCI research can guide the way in design ethics in practice research across other fields. In this discussion section, through our key findings and their implications, we synthesize a characterization of the turn to practice in design ethics research in terms of its approach to the concept of ethics, its shared motivations, research purposes and methods,  and suggest future directions for advancing this field.

\subsection{Characteristics of Design Ethics Research in Practice}

\subsubsection{Shared motivations:} Research on design ethics in HCI is increasingly driven by a sense of urgency to ensure that the growing power and influence of design are exercised ethically and responsibly. Many studies are turning to practice due to a growing recognition that ethical considerations in design are inherently situated and shaped by the complex realities of professional work \cite{chivukula_dimensions_2020, lindberg_design_2021, lindberg_cultivating_2022, jacobs_bridging_2021, spiel_details_2020}.  As a result, there is a marked shift toward emphasizing design ethics in HCI research. This focus also responds to a perceived disconnect between theoretical discussions of design ethics and the lived experiences of practitioners. Researchers seek to bridge this gap by engaging directly with practice, aiming to resonate with the conditions and challenges designers face in their everyday work \cite{gray_scaffolding_2023, lindberg_design_2021, dindler_engagements_2022, shilton_blended_2017}.

\subsubsection{Shared conceptual attributes of ethics:}Research that turns to practice to study design ethics discusses the concept of ethics as an intrinsic and evolving aspect of design work \cite{frauenberger_-action_2016}. Rather than treating ethics as a fixed or abstract set of principles, this approach situates ethical practice within the everyday realities of design—ranging from individual actions to organizational and societal dynamics \cite{popova_who_2024,gray_building_2024}. Ethics, in this view, is something to be actively cultivated through engagement with context-specific challenges and collaborative practices  \cite{lindberg_cultivating_2020}. This perspective opens space for designers to foster cultures of ethical reflection and responsibility as part of their work. Although the studies reviewed employ diverse conceptual frameworks to approach and define ethics, they generally converge around four common attributes of how ethics is conceptualized. Ethics is viewed as inherent in the design process (e.g., \cite{dindler_engagements_2022}), context-dependent (e.g., \cite{gray_ethical_2019}), continuously evolving (e.g., \cite{frauenberger_-action_2016}), and capable of being cultivated through reflective and collaborative engagement (e.g., \cite{lindberg_cultivating_2020}). These shared features form a conceptual foundation for how ethics is understood and enacted within design practice.

\subsubsection{Shared research purposes:} The research reflects a range of overlapping purposes. These include understanding how ethics unfolds in the everyday work of designers, supporting practitioners as they navigate ethical challenges, operationalizing ethical considerations within both design processes and organizational settings, and cultivating foundational cultures of ethical awareness and reflection (see Table \ref{tab:purpose}). Our review shows that design ethics in practice remains an emerging area of inquiry, as most research currently focuses on understanding ethics in context, with tools primarily developed for individuals or small teams. These studies offer valuable insights into the nuanced and complex nature of ethics as it plays out in real-world settings. In addition, many researchers develop tools and methods based on their empirical findings to support practitioners in addressing ethical challenges (e.g.,  \cite{ lujan_escalante_dancing_2021, cooper_design_2024, lecomte_improving_2022, luria_letters_2022, wong_timelines_2021}). On the other hand, broader, more systemic efforts—such as embedding ethics into organizational infrastructures or fostering sustained cultures of ethical reflection—are currently less common \cite{lindberg_cultivating_2022, frauenberger_-action_2016, spiel_details_2020, chivukula_envisioning_2024}. These studies contribute to concept development by synthesizing insights from multiple empirical cases, representing important steps toward building shared concepts and a more cohesive foundation for the field of design ethics in practice.

\subsubsection{Shared research methods:}To address the practical, everyday nature of ethical concerns in design, researchers commonly use both observational and interventional methods. Observational approaches—such as interviews and field studies—are employed to understand specific dimensions related to design ethics (e.g., \cite{karlsen_balancing_2022, jacobs_bridging_2021, petersen_empathetic_2017, chivukula_identity_2021, kelly_towards_2019,  beattie_its_2024,  millard_ethics_2024}). Interventional methods—including co-design and participatory design workshops—are used to support practitioners, operationalize ethics within projects, and cultivate ethical awareness in practice (e.g. \cite{lindberg_cultivating_2020, popova_voicing_2024, roosen_dialectical_2020, gram-hansen_participatory_2016}). Collectively, these perspectives point to a shift in design ethics research, from abstract principles to the messy realities of practice. This shift necessitates a range of methodologies, a focus on designerly approaches, and an acknowledgment of the emotional and contextual factors that influence ethical decision-making in design. 

All in all, with some research already explicitly positioning itself within the turn to practice in HCI \cite{frauenberger_-action_2016, dindler_engagements_2022, malinverni_autoethnographic_2016, lindberg_cultivating_2020, gray_ethical_2019,gray_dark_2018}, practice-oriented design ethics research shares many of the same characteristics as this broader shift.  Following the turn to practice in HCI, this strand of design ethics research (1) represents a move from abstract, isolated perspectives to an embedded, practitioner-centered view of real-world work, (2) emphasizes understanding situated practices, acknowledging that interaction and ethics are not standalone concerns but are, embedded in the complex realities of design and technology work, and (3) aims to bridge the research-practice gap by engaging directly with practitioners, valuing their knowledge, and addressing practical needs through context-sensitive methods like ethnography, co-design, and participatory inquiry. Thus, what we characterize as the turn to practice in design ethics exemplifies this shift by studying how ethical concerns emerge and are navigated within everyday design contexts, highlighting the importance of relevance, situatedness, and practitioner engagement in both ethical reflection and research methodology.

\subsection{Future Directions}
In this section, we will identify key gaps and potentials that emerged from our findings and propose six future directions that need to be addressed to advance research on design ethics in practice. Each proposed direction will be anchored in existing HCI literature, providing a deeper exploration of these gaps and outlining how they can evolve and contribute to future research in this area.

\subsubsection{Need for concept-building empirical research:}  As illustrated in Table \ref{tab:purpose} and Table \ref{tab:contributions}, the majority of research (60\%) in our pool aims to explore the current landscape of ethics in practice. Research on practice-oriented ethics is growing; however, with notable exceptions (e.g., \cite{frauenberger_-action_2016, gray_ethical_2019}), concept-building through empirical research remains limited.  Although the number of studies is increasing, design ethics in practice remains an emerging field. We believe it is timely to advocate for more context-aware, concept-building research grounded in lived experience—research that contributes to a shared conceptual foundation for practice-oriented design ethics by reflecting on common concerns and interests— (1) to bridge the theory-practice gap in design ethics, as one of the primary motivations for researchers to turn to practice \cite{lindberg_design_2021, lindberg_cultivating_2022, jacobs_bridging_2021}, and (2) to ensure that a rapidly growing field does not lose direction without clear purposes or standards to evaluate its value and contribution \cite{rogers_hci_2012, velloso_theorising_2025}. Our guiding question for this future direction is: What would concepts, frameworks, and theories in design ethics look like if they were meant to resonate with, and emerge from, everyday work practice?

The theory-practice gap has long been a point of discussion in broader HCI research \cite{rogers_hci_2012, velloso_theorising_2025}, with numerous contributions aimed at addressing this divide—contributions that design ethics research can benefit from. We suggest building on these existing discussions in HCI to initiate a more focused conversation within ethics-related research. This gap arises from the abstract nature of theory, the fast-paced, pragmatic demands of design work, limited communication between researchers and practitioners, and the complexity or inaccessibility of certain theoretical frameworks \cite{rogers_hci_2012, dalsgaard_between_2014}. Additionally, it has been argued that HCI’s interdisciplinary nature, coupled with differing evaluation criteria in academia and practice, further exacerbates this gap \cite{rogers_hci_2012, velloso_theorising_2025}. To address these challenges, researchers have proposed various bridging strategies. 

One common strategy, as Rogers points out \cite{rogers_hci_2012}, is to "import" theories from other fields to build correlations and make sense of design situations. An example from our review is that drawing on Activity Theory,  Gray et al. develop the concept of ethical mediators \cite{gray_ethical_2019}, which are factors that shape a designer’s ethical awareness and decision-making in UX practice. Ethical mediators help explain the ethical complexity of design by demonstrating how various influences continuously shape ethical engagement. Another strategy Rogers suggests is the development of “wild theories” \cite{rogers_hci_2012}— theories that emerge directly from messy, real-world contexts, addressing the interdependencies between design, technology, and behavior. These theories in HCI serve multiple purposes—not only to predict or prescribe, but also to generate ideas and sensitize designers to critical issues \cite{rogers_hci_2012, velloso_theorising_2025}. For example, Oulasvirta and Hornbæk \cite{oulasvista_counter} postulate that theories inform design practices through counterfactual thinking, enabling designers to speculate about the potential consequences of different design choices. This approach, when applied to concept development through empirical, practice-based research, can help bridge the gap between theory and practice in design ethics research.

In response to the need for more practically resonate approaches, many recent HCI efforts have focused on creating intermediary-level concepts instead of developing formal, scientific theories \cite{rogers_new_theo, velloso_theorising_2025}. For example, Höök and Löwgren \cite{hook_strong_2012} introduced strong concepts, which are abstracted from specific design cases and are meant to enrich designers’ repertoires. Though they may be supported by theory, they primarily emerge from practice. Another proposal comes from Dalsgaard and Dindler \cite{dalsgaard_between_2014}, who introduce bridging concepts—a form of intermediary knowledge that explicitly connects abstract theory with practical design. Unlike strong concepts, which primarily emerge from recurring design patterns, bridging concepts are grounded in both theoretical frameworks and real-world design examples. Their primary purpose is to facilitate a two-way exchange between theory and practice, holding both accountable, and helping to surface new design opportunities and theoretical insights. Such efforts emerging from HCI research can help guide design ethics research by fostering shared concepts and contributing to a more unified research landscape, while also helping to bridge the gap between theory and practice. 

\subsubsection{Need for longitudinal studies:} The sources we reviewed showcase a wide range of methodologies used to investigate design ethics in practice, with interviews being the dominant form of inquiry. Many studies utilize designerly methods and practices to address ethical challenges, such as design tools and techniques to explore ethical dilemmas, develop strategies for ethical intervention, and facilitate discussions about values and responsibilities in design. However, one clear gap we identified is the need for long-term studies; most research in the field relies on cross-sectional study designs. Interviews, while effective for capturing specific moments in time, often fall short of fully capturing situated practices, and are less effective in exploring processes and changes over time \cite{jarrahi_digital, chapman_social}. Since the ethics in practice approach views ethics as dynamic rather than static, long-term studies are essential to capture the evolving and temporal nature of ethics in design practice. Longitudinal studies can allow researchers to observe how ethical practices develop, how designers' ethical awareness grows, and how external factors influence ethical decisions. 

Longitudinal research in HCI involves collecting data from the same participants at multiple points in time, allowing researchers to study how user experiences, behaviors, and perceptions evolve \cite{karapanos_advances_2021}. While valuable, long-term studies present several challenges. They are time-consuming and costly, requiring sustained participant engagement, careful planning, advanced training, and often face methodological issues such as inconsistent data, small sample sizes, and complex analysis, which can make achieving valid results difficult \cite{harbich_user_2016, karapanos_advances_2021}. Despite notable shift in HCI toward studying temporality more explicitly as a central aspect of research \cite{wiberg_time_2021}, similar to design ethics research, it is stated that there remains a need for more longitudinal studies in HCI in general, partly due to a lack of shared understanding and clear definitions within the HCI research \cite{bargas-avila_old_2011, harbich_user_2016, karapanos_advances_2021}. To address this, Kjærup et al. \cite{kjearup_review_2021} note that researchers have been developing taxonomies to categorize types of longitudinal research questions and designs. Conferences like CHI have hosted workshops and discussions to build consensus on best practices, and literature reviews are examining how previous work has approached longitudinal study design (e.g. \cite{jain_workshop, karapona_workshop}). These efforts to create a common understanding can also guide ethics research.

Additionally, we believe sharing behind-the-scenes insights into design ethics could be valuable. This is an evolving area where researchers often deal with sensitive subjects and contested issues. Many researchers collaborate with design practitioners in commercial settings, where accessibility barriers, such as time constraints, economic pressures, and confidentiality concerns, are prevalent. By offering self-reflective experiences, strategies for conducting research in commercial settings, working with practice rich in implicit knowledge, and addressing potentially triggering issues, researchers can help guide others in conducting more long-term studies.

\subsubsection{Need for contextualizing design ethics research:} Design ethics research emphasizes the embedded nature of ethics, highlighting its situatedness within specific design practices, designer identities, organizational contexts, and broader socio-geographical conditions. These wider conditions shape the values and principles guiding both organizations and practitioners \cite{madaio_co-designing_2020, tekogul_cultivating_2024, lindberg_cultivating_2022, ozkaramanli_design_2022}. While ethics is widely recognized as a sociocultural construct that varies across cultures and geographies, there is a noticeable lack of research that fully contextualizes and localizes findings. Most research does not specify its location or relate its findings to the socio-geographical context in which the authors, participants, and stakeholders are situated. A notable exception is the work of Beattie et al. \cite{beattie_its_2024}, who contextualize the relationship between design practitioners and ethical decision-making within the broader framework of policies—or the lack thereof—in New Zealand. This underscores the need to reflect on how research findings align with the local nuances of where they are conducted, ensuring resonance with specific cultural and geographical contexts.

Understanding social context has become a central, though contested, concern in HCI \cite{rasanen_new_2006, rogers_hci_2012, joshi_who_2024}. Initially, context was often equated with "immediate" tasks and interactions, often reduced to observable actions, overlooking broader socio-cultural influences \cite{rasanen_new_2006}. This shifted with the adoption of ethnographic methods in the 1980s, which emphasized the complexities of users’ everyday work environments and the social dynamics shaping technology use \cite{joshi_who_2024}. Influential work, such as Lucy Suchman’s Situated Action theory, highlighted the improvisational and situated nature of interaction, challenging static views of context \cite{rogers_hci_2012}. However, debates persist over how context is treated as a mere setting or ethnography is used instrumentally \cite{rasanen_new_2006}. More recently, a cultural turn in HCI has expanded focus to include diverse areas like art and entertainment, stressing the importance of cultural values in design \cite{benford_ethical_2015} and first-person methods \cite{desjardins_introduction_2021}. Some examples can be exploration of rural islandness \cite{robinson_rural_2021}, anthropological conceptualization of hope in the context of Sub-Saharan Africa \cite{ratto_reopening_2023}, and the role of witchcraft in computing through villages of Bangladesh \cite{sultana_witchcraft_2019}.

However, it seems that cultural context often becomes a point of discussion only when research moves beyond urban, Western, and commonly studied contexts. In our own review, we focused exclusively on articles written in English. Partly because of this bias, research contexts are primarily based in Western countries, and the studies are typically conducted within researchers' own local settings, and socio-geographic and cultural contexts are merely mentioned. However, it’s important to recognize that Western contexts are not universal—they are specific cultural environments in their own right. Expanding HCI to include cultural perspectives is valuable, but it doesn't automatically address these complexities. Even within Western settings, when working in global organizations, understanding the local context remains important. This suggests there is room to further develop our awareness and sensitivity to cultural differences in research.

\subsubsection{Need for considering entanglements of design ethics:} Another point we want to bring into the discussion is the role of human and non-human entanglements in design ethics, building on Frauenberger’s explorations \cite{frauenberger_entanglement_2020} where they are inseparably linked. Verbeek \cite{verbeek_materializing_2006} explores the role of technological artifacts through the concept of technological mediation, highlighting how designed artifacts and society interact in ways that inherently involve ethical considerations. This aligns well with the turn to practice in design ethics, where every interaction within assemblages of human and non-human actors is framed as an ethical encounter, and with practice theory that takes artifacts as crucial actors within practices \cite{kuijer_practices_2013, entwistle_beyond_2015}. This idea extends to the everyday tools and software design practitioners use, which influence their ethical decision-making processes.

Frauenberger \cite{frauenberger_entanglement_2020} argues that the concept of entanglement may represent the next major paradigm in HCI, as the field faces increasing pressure from the evolving relationship between humans and digital technologies. He observes that the rapid pace of technological innovation and profound social transformations are exposing the limitations of HCI’s existing foundations such as inability to address ontological uncertainty, limitations in knowledge production, difficulties in assigning responsibility within complex socio-technical systems, reliance on a static concept of the user, and outdated ethical frameworks that fail to reflect the dynamic and relational nature of human-technology entanglements \cite{Fuchsberger_respon, frauenberger_entanglement_2020}. In the context of design ethics, the entanglement concept is suggested to imply a significant shift in how ethics \cite{frauenberger_entanglement_2020} and responsibilities \cite{Fuchsberger_respon} are understood—particularly as technologies gain autonomy and exert influence beyond direct human control. Fuchsberger and Frauenberger \cite{Fuchsberger_respon} highlight the resulting “responsibility gap,” where it becomes difficult to assign blame or moral accountability through traditional frameworks. They address this challenge by emphasizing that responsibility emerges through ongoing interactions among human and nonhuman actors across all stages of technological development. For design practice, this perspective means treating technology as an active participant, focusing on configuring relationships rather than isolated actions, and adopting inclusive, participatory methods that recognize the complex interplay of actors involved in shaping technological futures \cite{Fuchsberger_respon}.

In our review, we observe that the subjects included in the studies we reviewed, predominantly practitioners, function as human actors within the design process. While our review includes research on tools and methods specifically designed to support ethical decision-making (e.g., \cite{gray_using_2022}), there is a lack of studies addressing the interconnections between everyday design tools, software, and practitioners' ethical decision-making. Empirically, we suggest that this approach encourages us to foreground the materialities of the design process as research subjects. It highlights the active role these tools play as ethical mediators and actors within the process. Exploring this perspective further could inform the development of design tools that align more closely with the ethical and practical dimensions of the design process.

\subsubsection{Need for specialized design methods for cultivating ethics:} This review documents the interventional methods employed in design ethics research. The interventional quality of these methods is distinctive to the way ethics is approached from a practice-oriented perspective. Within this perspective, ethics is not seen as a fixed set of principles but as something that is evolving, collaboratively constructed, and actively cultivated through practice. This framing of ethics calls for methodological approaches that go beyond traditional qualitative tools. Designerly methods—particularly speculative and co-design approaches—are especially valuable in this context.

There is already a growing body of interventional methods developed by HCI and design researchers that contribute meaningfully to ethics research, particularly in fostering a collaborative ethos and cultivating shared ethical understanding. Speculative methods and design fiction, for instance, have long served as tools for critical reflection on emerging technologies. These methods make use of imaginative scenarios and designed artifacts to probe ethical tensions, stimulate discussion, engage diverse perspectives, and encourage critical reflection on the values that underpin technological development \cite{haghighi_workshop-based_2023, wong_timelines_2021, tekogul_cultivating_2024}. Similarly, co-design and participatory design emphasize the inherently collaborative and situated nature of ethical inquiry. They provide interactive formats for exploring ethical concerns with stakeholders, reinforcing the view of ethics as cumulative, contextual, and socially negotiated \cite{lindberg_cultivating_2020, lindberg_cultivating_2022}. We argue that HCI and design research are well-positioned to develop specialized methods for this context, drawing on their foundations in speculative and participatory approaches. We argue that HCI and design research are uniquely positioned to advance this area by developing specialized methods for ethics research, building on their strengths in speculative and participatory traditions. These approaches provide the conceptual and practical tools needed to support collaborative, reflective, and practice-based engagement with ethical challenges in design.

\subsubsection{Need for making ethics engaging:} A general observation we had is that the language surrounding ethical issues often carries negative connotations, such as avoiding deceptive designs, minimizing risks, and avoiding harmful future implications. Instead of framing ethics as the avoidance of unfavorable outcomes, how can we position it as something engaging and approachable? We find traces of this question in the review; for instance, Lindberg et al. \cite{lindberg_cultivating_2022}  mention the importance of “presenting ethics in an appealing manner”, and Escalante et al. \cite{lujan_escalante_dancing_2021} position ethics as a capacity rather than a duty. This is an open question that we propose for future research to explore in a more structured way.

\section{Conclusion}
We have presented a scoping review of practice-oriented research in design ethics, characterizing the design fields’ expansion in relation to the turn to practice in HCI research. Our review highlights a growing interest in understanding and addressing the ethical implications of design, emphasizing the value of practice-oriented approaches in capturing the situated, tacit, and complex nature of design ethics. It also demonstrates the potential of such approaches to develop concepts, tools, and methods that better align with the realities of design practice.

The review findings highlight several future directions that need to be addressed to advance research on design ethics in practice. While empirical studies in this area are increasing, there remains a significant lack of concept-building empirical work. Current research predominantly relies on cross-sectional study designs with interviews as the primary method, highlighting the need for long-term studies to enhance understanding and address the temporality of ethics. Design ethics are recognized as socially and culturally situated, but further research is needed to socio-geographically contextualize these studies. Although existing research explores specialized tools and methods for ethical consideration in the design process, there is a need to examine how existing design tools mediate ethical practices. This review also emphasizes the significance of design methods that can bring collaborative approaches to ethics cultivation and position ethics as an active, engaging, and resourceful capacity in design, rather than approaching it with caution or hesitation. Finally, we note that design ethics research spans a broad spectrum of fields, with the design in HCI research well-positioned to guide design ethics in practice studies across disciplines, with its cumulative experience in practice-oriented research for grappling with various impacts of technology development. 


\begin{acks}
This work is funded by the Independent Research Fund Denmark under Grant No. 3097-00058B.
\end{acks}
\bibliographystyle{ACM-Reference-Format}


\end{document}